\documentclass[11pt,a4paper]{article}

\RequirePackage[l2tabu, orthodox]{nag}

\usepackage{jheppub}
\usepackage{amsthm,amssymb,amsmath,epic,float}
\usepackage{rotating,epsfig,indentfirst,array,varioref}
\usepackage{appendix,tikz,mathtools}
\usepackage{setspace}
\usepackage{tikz}
\usetikzlibrary{decorations.pathreplacing}
\usetikzlibrary{calc}
\usepackage{enumitem}
\usepackage{hyphenat}
\usetikzlibrary{positioning}
\usepackage{graphicx}  
\usepackage{adjustbox}

\def\be{\begin{equation}}
\def\ee{\end{equation}}

\usepackage{jheppub}
\makeatletter
\def\@fpheader{\vspace{-.1cm}}
\makeatother

\title{\boldmath  Periods of the multiple Berglund–Hübsch–Krawitz mirrors  }

\author[a,c,d]{Alexander Belavin,}
\author[b]{Vladimir Belavin,}
\author [d]{Gleb Koshevoy}

\affiliation[a]{Landau Institute for Theoretical Physics, 142432 Chernogolovka, Russia}
\affiliation[b]{Physics Department, Ariel University, Ariel 40700, Israel}
\affiliation[c]{Moscow Institute of Physics and Technology, 141700 Dolgoprudny, Russia}
\affiliation[d]{Kharkevich Institute for Information Transmission Problems, 127994
Moscow, Russia}

\emailAdd{belavin@itp.ac.ru}
\emailAdd{vladimirbe@ariel.ac.il}

\abstract{We consider the multiple Calaby-Yau (CY) mirror phenomenon which appears in  Berglund-Hübsch-Krawitz (BHK) mirror symmetry. 
We show that for any pair of Calabi--Yau orbifolds that are BHK mirrors
of a loop--chain type pair of Calabi--Yau manifolds in the same weighted
projective space the periods of the holomorphic nonvanishing form coincide.}

\dedicated{In memory of Boris Dubrovin.}


\begin{document} 
\maketitle
\flushbottom

\section{Introduction}
\label{sec:intro}

Calabi-Yau manifolds (CY) arise in the context of spacetime supersymmetric compactifications of string theories. An important property of CY manifolds is  mirror symmetry 
\cite{Gepner,Greene, Hori}, which reflects geometric relation between pairs of CY manifolds. 
Namely, for a pair of n-dimensional Calabi-Yau manifolds  $X$ and $Y$,  cohomologies are  isomorphic, $ H^{p,q}(X,\mathbb{C})=H^{n-p,q}(Y,\mathbb{C})$.
Calabi-Yau manifolds possess the structure of complex and Kähler manifold which admit deformations (see~\cite{Candelas:1990pi}), giving rise to the moduli spaces $M_C(X)$  and $M_K(X)$. 
Mirror symmetry can be considered as matching of the Special geometries \cite{Strominger} on the moduli spaces  
\begin{equation}
    M_C(X) \simeq M_K(Y)\;, \quad M_K(X) \simeq M_C(Y)\;.
\end{equation}
Here we focus on the class of CY manifolds, defined as hypersurfaces or orbifolds in weighted projective spaces $\mathbb{P}^4_{(k_1,k_2,k_3,k_4,k_5)}$  cut out  by   non-degenerate quasihomogeneous polynomials consisting of five monomials. 
The nondegeneracy requires that these polynomials must be a sum of the polynomials of  three basic types, classified  by M.~Kreuzer and 
H.~Skarke \cite{Kreuzer}.
For this class of CY manifolds it was suggested by Berglund and Hübsch  in
\cite {Berglund}  that  their mirrors are obtained  as particular orbifolds of the projective spaces defined by the ``transposed'' polynomials. The Berglund-Hübsch  mirror construction was generalized by Krawitz in \cite{Krawitz}.
 Chiodo and Ruan \cite{Chiodo}  proved that the Berglund-H\"ubsch-Krawitz (BHK) mirrors form a mirror pair on the level of Chen-Ruan orbifold cohomology.

The multiple mirror  phenomenon occurs when a given CY threefold possesses  more than one  mirror in different weighted projective spaces.
More specifically the phenomenon is the following.  Some  weighted projective spaces allow the existence of a few CY manifolds defined by the polynomials that belong to  different types of Kreuzer-Skarke class.
In such cases the Berglund-Hübsch-Krawitz (BHK) mirrors of these two CY manifolds generally  appear in two  different weighted projective spaces.
The examples of such phenomenon are given in the Table \ref{table:Chain-Loop-Mirrors}, which   shows all the cases when  the weighted  projective space allow the  simultaneous occurrence of CY-manifolds of loop and chain types of Kreuzer-Skarke class CY manifolds.
The problem of birationality  of multiple BHK-mirrors has been investigated early in  the case of one-parameter CY family \cite{Kelly, Shoemaker, Clarke, Clader},  
where it was shown that the different BHK mirrors are birational.
The derived equivalence of such multiple BHK-mirrors was obtained in 
\cite{Favero}.
In \cite{ABMB}, the periods of the multiple mirrors were calculated in one case of multi-parameter CY family, and it was found that they match.
In this paper we prove  that the periods of the nonvanishing holomorphic form 
$\Omega$ coincide for multiple BHK  mirrors of the  loop and chain types.
We assume that, using a similar technique,  one can also  prove 
   coincidence of the periods  for other types of multiple BHK mirrors.

In \cite{G} it was shown  using the technique of Griffiths-Dwork diagrams, that if two invertible pencils 
($\sum_j \prod_ix_ {ij}^{M_ {ij}} + t \prod_ix_i $) with reversible
$M$ have the same dual weights, then their Picard-Fuchs equations are the same.  
It is equivalent to the statement that one-parameter deformations of several BHK mirrors have the same Picard-Fuchs equations. This statement is 
consistent with ours as a special case.
The paper \cite{G} used another tools to show the coincidence of Picard-Fucks equations for one parameter deformation of invertible polynomials with multiple mirrors. Note that the coincidence of the periods does not  follow automatically from this even in this particular case. 

In Section~\ref{sec:2} we briefly  recall the BHK mirror construction.
In Section \ref{sec:3}, we consider the expressions for the periods of the holomorphic form $\Omega$ for   two different BHK mirrors  and show their coincidence.

\section{Berglund–Hübsch–Krawitz mirror construction}
\label{sec:2}

As was mentioned above among all possible  CY manifolds there is a class \cite{Kreuzer}, which is defined  in the weighted projective space
\begin{equation}
\label{weighted}
\mathbb{P}^4_{(k_1,k_2,k_3,k_4,k_5)}=\Big\{ (x_1,\dots,x_5)\in \mathbb{C}^5 \setminus \{0\} \ \Big| \ x_i \sim \lambda^{k_i} x_i \Big\}
\end{equation}  
by a quasi-homogeneous polynomial  
\begin{equation}
\label{eq:CY}    
W^0_M(x)=\sum_{i=1}^5 \prod_{j=1}^5 x_j^{M_{ij}}\;.
\end{equation}
Such polynomials obey the following properties:
\begin{itemize}
\item[1)] Matrix $M$ is integer, and invertible;
\item[2)] The polynomial $W^0_M$ is quasi-homogeneous, i.e.,
 there exist positive integers $k_i$ (weights of the 
projective space), and $d$ (degree), such that
\begin{equation}
\label{eq:CYcondition}
    \sum_{j=1}^5 M_{ij}k_j=d, \ \ \ \forall i,
\end{equation}
where $d=\sum_{j=1}^5k_j$, which is CY condition for the orbifold in the wighted projective space. 

\item[3)] The polynomial $W^0_M$ is non-degenerate outside the origin. 
\end{itemize}

From these conditions it follows,  that the $W^0_M$ is given by a sum of polynomials 
 of one of three  types~\cite{Kreuzer}: 
\be\label{A_i}
\begin{aligned}
&x_1^{A_1}+x_2^{A_2}+...+x_n^{A_n} - \text{Fermat}\;,\\
&x_1^{A_1} x_2 + x_2^{A_2} x_3 +...+x_n^{A_n} - \text{Chain}\;,\\
&x_1^{A_1} x_2 + x_2^{A_2} x_3 + ...+x_n^{A_n}x_1 - \text{Loop}\;.
\end{aligned}
\ee

Then it follows  that the equation  $W^0_M=0$ defines a Calabi-Yau orbifold $X^0_M$.  
We use below the following  properties of the matrix $M$.
There  exist positive integer numbers $\bar{k}_j$ such that 
\begin{equation}
\label{eq:CYconditionT}
   \sum_{i=1}^5\bar{k}_i M_{ij}=\bar{d}, \ \ \ \forall j,
\end{equation}
where $\bar{d}=\sum_{i=1}^5 \bar{k}_i$.
The set $\bar{k}_j$ is nothing but the set of weights of  another weighted projective space
 $\mathbb{P}^4_{(\bar{k}_1,\bar{k}_2,\bar{k}_3,\bar{k}_4,\bar{k}_5)}$ where $ X_{M^T}$, i.e. the mirror
of the CY orbifold $X^0_M$, which will be defined below. Let $B_{ij}$ denote the inverse matrix of $M$, i.e.  $\sum_k B_{ik}M_{kj}=\delta_{ij}$. Then using (\ref{eq:CYcondition}),(\ref{eq:CYconditionT})  
 we get the relations (see e.g.~\cite{Chiodo}):
  \begin{equation}
\label{eq:matrixB-BT}
    \sum_{j=1}^5 B_{ij}=\frac{k_i}{d}\;,\quad  \sum_{i=1}^5B_{ij}=\frac{\bar{k}_j}{\bar{d}}\;.
\end{equation}

The CY orbifold $X^0_M$ admits  a deformation of the complex structure, which  is realized by a polynomial $W_M$, which is  a deformation of the $W^0_M$. 
The full family  $X_M$ is given by zero locus of 
\begin{equation}
\label{eq:fullfamillyofX}
    W_M=\sum_{i=1}^5 \prod_{j=1}^5 x_j^{M_{ij}}+\sum_{l=1}^{h}\varphi_l e_l(x),
\end{equation}
where  $e_l$ are some quasi-homogeneous monomials (see below). 
 The $\varphi_l$ are moduli of the complex structure of the family $X_M$, 
and $h:=h_{21}$ is the Hodge number of the family $X_M$, which is equal to the dimension of the 
complex structures moduli space of $X_M$.

Berglund and Hübsch proposed \cite{Berglund} 
that the mirror partner  $X_{M^T}^0$ of the orbifold $X_M^0$ is defined
as a quotient of the zero-locus of the polynomial 
\begin{equation}
    W^0_{M^T}(z)=\sum_{i=1}^5\prod_{j=1}^5 x_j^{(M^T)_{ij}}
\end{equation}
 in the weighted projective space $\mathbb{P}^4_{(\bar{k}_1,\bar{k}_2,\bar{k}_3,\bar{k}_4,\bar{k}_5)}$. This quotient is realized by some subgroup of the phase symmetry group  of $W^0_{M^T}$.

This approach has been generalized by Krawitz \cite{Krawitz} as follows.
Let $\text{Aut}(M)$ be the group of the  phase symmetries
of $W^0_M=0$, and  let $\operatorname{SL}(M)$ be its maximal subgroup,
which preserves the nonvanishing holomorphic form  $\Omega$. 
We will call $\operatorname{SL}(M)$  the maximal admissible group.
We denote by $J_M \subseteq \operatorname{SL}(M)$  the subgroup   that consists of the phase symmetries induced by $\mathbb{C}^*$ action on  $\mathbb{P}^4_{(k_1,k_2,k_3,k_4,k_5)}$.
Choose a group $G_0$ to be some subgroup of $\operatorname{SL}(M)$ containing $J_M$ and let  $G=G_0/J_M$.
 It can be  shown that the quotient space $Z(M,G):=X_M/G$ is a CY orbifold.  
The same can be applied to the transposed polynomial $W_{M^T}^0$
 to get CY orbifold $Z(M^T,G^T):=X_{M^T}/G^T$. 
Krawitz has shown how to choose the dual group $G^T$ such that 
$Z(M,G)$ and $Z(M^T,G^T)$ form a mirror pair. 
Below we remind this mirror constuction ~\cite{Krawitz} in more details.

 Let CY hypersurface $X_M^0$ be defined, as above,  
by the zero locus of $W_M^0(x)$ and  let $\omega= e^{2\pi i/d}$. 
The polynomial  $W_M^0(x)$, as well as the form  $\Omega$,  is invariant  under the action of the group $J_M$   defined as
\begin{equation}
\label{eq:QMaction}
    x_i\mapsto \omega^{k_i} x_i, \ \ \ \omega^d=1.
\end{equation}
 Let  $q_i(M)$ be  generators of ${Aut}(M)$, which act on the coordinates as
\begin{equation}
    q_i(M): x_j \mapsto e^{2\pi i B_{ji}} x_j.
\end{equation}
In these terms, the generator of the group $J_M$ 
\begin{equation}
   \hat{J}= \prod_{i=1}^5q_i(M).
\end{equation} 
 Let  $G_0$ be some admissible group such 
that $J_M \subseteq G_0 \subseteq SL(M)$ and  the quotient group
\begin{equation}
    G:=G_0/J_M.
\end{equation} 
Then the orbifold defined as $Z(M,G):=X_M/G$ is a Calabi-Yau manifold.
Note that $X_M$ itself is a special case of $Z(M,G)$ when $G_0=J_M$.

The BHK  construction begins with the  polynomial $W_M$ and some group $G_0$. 
 Let  monomials
\begin{equation}
    \label{eq:Monomials} 
    e_l:=\prod_{j=1}^5 x_j^{S_{lj}},
\end{equation}
where $\sum_{j}S_{lj}k_j=d$,
be elements of the $G_0$-invariant subring of the Milnor  ring   
$\mathbb{C}[x_1,\dots,x_5]/ \langle  \frac{\partial W_M}{\partial x_j} \rangle$ \cite{Lerche}.
Then the quotient by the group $G$ is defined by the zero locus of the polynomial
\begin{equation}
\label{eq:Familly}
    W_M=\sum_{i=1}^5 \prod_{j=1}^5 x_j^{M_{ij}}+\sum_{l=1}^{h}\varphi_l \prod_{j=1}^5 x_j^{S_{lj}},
\end{equation}
where $\varphi_l$ are moduli of complex structure deformations, and $h:=h_{21}(X)$ is Hodge number.\footnote{With a slight abuse of notation, we use here $h$ for the Hodge number of the  orbifold, which we used before for the original manifold $X_M$.}
 We denote by $e_{h}$ the monomial 
with $S_{h,i}=1$, which  plays a distinguished role, 
\begin{equation}
\label{eq:eh}
    e_{h}=\prod_{i=1}^5x_i.
\end{equation}


	Let $\rho_s$ be the generators of the group $G_0$, then by definition
\begin{equation}
    \rho_s \cdot \prod_{j=1}^5 x_j^{S_{lj}} = \prod_{j=1}^5 x_j^{S_{lj}} , \ \ \ l=1,\cdots,h.
\end{equation} 
We can also define the similar data for the  polynomial  $W_M^T$ and for some of its admissable group $G^T_0$.
 Taking the quotient $G^T:=G_0^T/J_{M^T}$, we get the Calabi-Yau orbifold
$Z(M^T,G^T)$.

  The way to choose  a group $G^T$  such that the manifold $Z(M^T,G^T)$  will be the mirror for $Z(M,G)$ is as follows~\cite{Krawitz}.
The generators $\rho_l^T$ of the group $G_0^T$ should be  constructed using the exponents $S_{li}$ of the invariant monomials 
 (\ref{eq:Monomials}) as follows
\begin{equation}
    \rho_l^T:=\prod_{i=1}^5 q_i(M^T)^{S_{li}},
\end{equation}
where $q_i(M^T)$ are the generators of $\operatorname{Aut}(M^T)$ acting on each coordinate $z_j $  in $\mathbb{P}^4_{(\bar{k}_1,\bar{k}_2,\bar{k}_3,\bar{k}_4,\bar{k}_5)}$ as
\begin{equation}
    q_i(M^T): z_j \mapsto e^{2\pi i B_{ji}^T}z_j=e^{2\pi i B_{ij}}z_j.
\end{equation}
It follows that the generators $\rho_l^T$ of  the group $G_0^T$ act on the coordinates $z_j $ as
\begin{equation}
\label{eq:HTaction}
    \rho_l^T: z_j \mapsto e^{2 \pi i \sum_i S_{li}B_{ij}} z_j.
\end{equation}
The  mirror Calabi-Yau family $Z(M^T,G^T)$ is given by the zero locus of 
\begin{equation}
\label{eq:CalabiYauMirror}
    W_{M^T}:= \sum_{i=1}^5 \prod_{j=1}^5  z_j^{M^T_{ij}}+ \sum_{r=1}^{\bar h} \psi_r \prod_{j=1}^5 z_j^{T_{rj}}
		= \sum_{i=1}^5 \prod_{j=1}^5  z_j^{M_{ji}}+ \sum_{r=1}^{\bar h} \psi_r \prod_{j=1}^5 z_j^{T_{rj}},
\end{equation}
where $\psi_r$ are moduli of the complex structure of the family $Z(M^T,G^T)$ and  monomials $\bar{e}_r:=\prod_{j=1}^5 z_j^{T_{rj}}$ are invariant under the $G_0^T$ action (\ref{eq:HTaction}):
\begin{equation}
\label{eq:invcond}
    \rho_l^T \cdot \bar{e}_r = e^{2 \pi i \sum_{ij} B_{ij}S_{li}T_{rj}} \prod_{j=1}^5 z_j^{T_{rj}} = \prod_{j=1}^5 z_j^{T_{rj}}.
\end{equation}
It is also necessary to take into account the quasi-homogeneity condition
\begin{equation}
    \label{eq:homogmirror}
    \sum_{i=1}^5T_{mi}\bar{k}_i=\bar{d}.
\end{equation}
It is convenient to consider the exponents $S_{li}$, and $T_{rj}$ as   nonnegative integer five-component  vectors $(\vec{S}_l)_j=S_{lj}$, 
and $(\vec{T}_r)_i=T_{ri}$. 
It was shown in \cite{ABBE} that  the condition (\ref{eq:invcond})  can be reformulated in terms of the pairing of these vectors, defined by the matrix $B$, as follows
\begin{equation}
\label{eq:pairing}
    (\vec{S}_l,\vec{T}_r):=\sum_{i,j=1}^5 B_{ij}S_{li}T_{rj} \in \mathbb{Z}.
\end{equation}
The relations (\ref{eq:homogmirror}) and (\ref{eq:pairing}) are very restrictive because  the entries of the matrices $S_{li}$ and $T_{rj}$ are integers, while the entries 
of the inverse matrix $B_{ij}$ are rational. Therefore,  if the vectors $\vec{S}_l$ are defined as above, these relations have a finite number of solutions for the vectors $\vec{T}_r$.
The CY orbifold cut out by $W_{M^T}$  is the mirror for $Z(M,G)$, as it is shown in~\cite{Krawitz}.


\section{Periods of BHK mirrors and their coincidence} 
\label{sec:3}

In this section we use the relation (\ref{eq:pairing}) to prove our basic statement that the periods of multiple mirrors coincide.
 We will focus on the case when two hypersurfaces 
in the weighted projective space simultaneously come from two polynomials of the loop and chain types that ``point'' in the same direction.
We will assume also that the admissible group $G_0=J_M$ preserves both polynomials. 

 Let $X_M$ be the CY family  $X_M$ as defined in Section 2.
 We regard $X(M)$  as the family of CY manifolds, since  the polynomial 
 $W_M(x,\phi)$ is a function of two sets of variables $x$ and parameters $\{\phi_l\}$. 
 Then  the mirror family $Y(M^T)$ is defined in 
$\mathbb{P}_{\vec{\bar{k}}}=
\mathbb{P}^4_{(\bar{k}_1,\bar{k}_2,\bar{k}_3,\bar{k}_4,\bar{k}_5)}$  by zeros of the polynomial 
 \be\label{mirrorY}
 W_{M^T}(x) =\sum_{i=1}^5 \prod_{j=1}^5 x_j^{M^T_{ij}}+ 
\sum_{r=1}^{\bar{h}} \psi_r \prod_{j=1}^5 x_j^{T_{rj}}\;,
 \ee
  where it is assumed that  the relationships 
$\sum_{j=1}^5 T_{rj} \bar{k}_j =\bar{d}$ and 
$\sum_{ij}B_{ij} S_{li} T_{rj} \in \mathbb{Z}$ are fulfilled.

 Periods of the holomorphic form $\Omega$ for CY family $X(M)$ can be  rewritten as  the contour integrals \cite{ABKA}
 \be
 \sigma_{\vec{\mu}}(\phi)=\oint_{\Gamma_{\vec{\mu}}} e^{-W_M(x,\phi)} d^5 x\;. 
 \ee
 The number of the contours $\Gamma_{\vec{\mu}}$ is equal 
to $\text{dim}\, R(M)=2h+2$.
Here $R(M)$ is so-called chiral ring which is subring of polynomials in $x_1,...,x_5$   modulo 
$\{\frac{\partial W^0_M}{\partial x_i}\}$ of degree  $0$,$d$, $2d$ or $3d$.
 The ring $R(M)$ consists of monomials
 \be\label{basis}
 e_{\vec{\mu}}(x)=\prod_{i=1}^5 x_i^{\mu_i} \;,
 \ee
where
 \be\label{grade}
 \sum_{i=1}^5 k_i \mu_i =0,\,\,d,\,\,2d,\,\,3d 
 \ee
 and $\mu_i$  are non-negative integers  satisfying some constraints:\footnote{Below $A_i$ are defined according to eqs.\eqref{eq:CYcondition} and \eqref{A_i}.}\\ 
 In the Fermat case
 \be
 0\leq \mu_i\leq A_i-2\;.
 \ee
 In the Loop case
 \be\label{cond-Loop}
 0\leq \mu_i\leq A_{i}-1\;.
 \ee
 A convenient set of the chiral ring  monomials in the Chain case  
was described in \cite {Kreuzer3} (see also \cite{Kreuzer2}, \cite {Krawitz}).

From~\eqref{grade} it follows that the ring $R(M)$ is graded,
 \be
 R(M)= \mathbb{R}^0(M)\oplus  R^d(M)\oplus  R^{2d}(M) \oplus R^{3d}(M)
 \ee 
 and $\text{dim}\, R^0(M)=\text{dim}\,  R^{3d}(M)=1$, while 
$\text{dim}\,  R^d(M)=\text{dim}\, R^{2d}(M)=h$.
 The monomials $ e_l(x)=\prod_{j=1}^5 x_j^{S_{lj}},\, l=1,...,h$ are elements of  the subspace $R^d(M)$.
 Similar statements can be  formulated for the mirror orbifold $Y(M^T)$.\\
 Since
 \be
 W_M(x,\phi)=W_M^0(x)+\sum_{l=1}^h \phi_l e_l(x)\;, 
 \ee
 the periods have the form
 \be
 \begin{aligned}
 &\sigma_{\vec{\mu}}(\phi_1,...,\phi_h)=\int_{\Gamma_{\vec{\mu}}} d^5x e^{-W_M^0(x)} 
e^{-\sum_{l=1}^h \phi_l e_l(x)}
=\sum_{\{m_l\}}\prod_{l=1}^h  \frac{\phi_l^{m_l}}{m_l!}  \int_{\Gamma_{\vec{\mu}}} d^5x e^{-W_M^0(x)} \prod_{j=1}^5 x_j^{\sum_{l=1}^h m_l S_{lj}}.
 \end{aligned}
 \ee
Using the standard cohomology technique \cite{ABKA}, we obtain the following explicit expression 

\be
\sigma_{\vec{\mu}} (\phi_1,...,\phi_h)\sim \sum_{\{m_l\}\in \Sigma_{\vec{\mu}}} C(\{m_l\})
\prod_{l=1}^h \frac{\phi_l^{m_l}}{m_l!} \;, 
\ee
where
\be\label{period}
C(\{m_l\})=\prod_{i=1}^5 \Gamma\left(\sum_j(m_l S_{lj}+1)B_{ji}\right)\;.
\ee
The canonical choice of $\Gamma_{\vec{\mu}}$ defines it as
 a cycle dual to the element of the base of the chiral ring $e_{\vec{\mu}}$ 
\be
\int_{\Gamma_{\vec{\mu}}} e_{\vec{\nu}}(x) e^{-W_0(x)} d^5 x =
\delta_{\vec{\mu},\vec{\nu}}\;.
\ee
 It leads to the following definition of $\Sigma_{\vec{\mu}}$~\cite{ABKA}.
The set of non-negative integers 
$ m_l$, where $l=1,...,h$, belongs to  $\Sigma_{\vec{\mu}}$, if one can find such non-negative integers  $n_j$, $j=1,...,5$ that
\be
m_l S_{li}=({\vec{\mu}})_i+\sum_j n_{j} M_{ji}\;.
\ee
 From~\eqref{period} we see that the periods are determined  by the sets of $h$ 5-component vectors $\sum_{j=1}S_{lj}B_{ji}$, where $l=1,..,h$,  as well as by the sets $\{m_l\}$, $l=1,...,h$.

Let's now prove  that the periods of CY orbifolds 
$Y(M^T(1))$ and $Y(M^T(2))$, defined in two 
different  weighted  
projective spaces $\mathbb{P}_{\vec{k}(1)}$ and $\mathbb{P}_{\vec{k}(2)}$ are the same, because they are the BHK mirrors 
of the CY manifolds $X(M(1))$ and $X(M(2))$ located in the same projective space $\mathbb{P}_{\vec{k}}$. 

In order to simplify notations we denote the initial  loop matrix as $M(1)$, 
and the initial chain matrix as $M(2)$.
Recall that they satisfy  the relations
\be\label{M-k-d}
\sum_{j=1}^5 (M(1))_{ij}k_j =\sum_{j=1}^5 (M(2))_{ij} k_j=d=\sum_{j=1}^5 k_j\;.
\ee
We denote by $S_{lj}(1)\in \mathbb{Z}_{\geq 0}$ and $S_{lj}(2)\in \mathbb{Z}_{\geq 0}$ the  vectors of the exponents of the monomials providing the deformation of  the complex structure 
\be
e_l(1) =\prod_{j=1}^5 x_j^{S_{lj}(1)}\;,\quad e_l(2) =\prod_{j=1}^5 x_j^{S_{lj}(2)}
\ee
and
\be\label{S(1,2)}
\sum_{j=1}^5 S_{lj}(1) k_j=\sum_{j=1}^5 S_{lj}(2) k_j=d\;.
\ee 
The sets of vectors $\vec{S}_l(1)$ and $\vec{S}_l(2)$ are slightly  different. It follows from the  conditions 
for the sets of the chiral ring monomials in the Loop and Chain cases and also from the fact
that $A_i(1)=A_i(2)$ for $i=1,2,3,4$ and $A_5(1) \neq A_5(2)$ in the considered case.
  
 However, from Table 1 in the appendix we come to the following important observation:
{\it When a loop polynomial $W_{M{(1)}}$ and a chain polynomial $W_{M{(2)}}$ appear in the same weighted projective space 
$\mathbb P(k_1, \ldots , k_5)$ and are connected by Kreuzer-Skarke cleaves~\cite{Favero}, in all these 111 cases listed in Table 1  the weight $k_5=1$.}

 From this fact  we obtain that the sets of  exponents of  monomials of their chiral rings  shifted   by the vector $(1,\ldots ,1)$
 not only belong to the same  4-dimensial lattice, but also 
both contain the integral basis of the lattice, which consists of the four vectors $\vec{V_a},a=1,2,3,4$
\[
\begin{matrix}
(-1, 0, 0, 0, k_1),\cr
(0, -1, 0, 0, k_2),\cr
(0,  0,-1, 0, k_3),\cr
(0,  0,  0,-1, k_{4}).\cr
\end{matrix}
\]
Below we will use this observation   to prove our main statement.

Recall that the mirrors $Y(1)$ and $Y(2)$ are CY orbifolds in the weighted projective spaces $P_{\vec{k}(1)}$ and $P_{\vec{k}(2)}$ respectively.
The  weights ${\vec{k}(1)}$ and  ${\vec{k}(2)}$ satisfy the equations
\be\label{M-k-d-1-2}
\sum_{j=1}^5 (M^T(\alpha))_{ij}\bar{k}_j(\alpha) =\bar{d}(\alpha)=\sum_{j=1}^5 \bar{k}_j(\alpha)\;, \quad \alpha=1,2\;.
\ee
The two sets of the vectors $T_{mj}(1)$ and  $T_{mj}(2)$, defining the complex structure deformations, also satisfy the equations 
\be
\sum_{j=1}^5 (T(\alpha))_{rj}\bar{k}_j(\alpha) =\bar{d}(\alpha)\;, \quad \alpha=1,2\;.
\ee
From the BHK mirror conditions~\eqref{eq:pairing} it follows
\be
(\vec{S}(\alpha)_l,\vec{T}(\alpha)_r)=\sum_{i,j=1}^5 
B(\alpha)_{ij}S(\alpha)_{li}T(\alpha)_{rj} \in \mathbb{Z}\;, \quad \alpha=1,2\;,
\ee
where as before $B(\alpha)$ is the inverse matrix for $M (\alpha)$ 
and $B^T(\alpha)$ is its transpose.
From these equations and from the fact that two sets of vectors $\vec{S}_l(1)$ and 
$\vec{S}_l(2)$ are linear combinations of the same  basis vectors $\vec{V_a}$ with integer coefficients it follows that
\be\label{R-B}
T_{ri}(1) B^T_{ij}(1) =T_{ri}(2) B^T_{ij}(2)\;.
\ee


This equality  implies also the equivalence of the pairings
\be\label{SR}
(\vec{S}(1)_l,\vec{T}(1)_r)=(\vec{S}(2)_l,\vec{T}(2)_r)\;,
\ee
which in what follows for briefness is denoted $(\vec{S}_l,\vec{T}_r)$.

For the periods of the two BHK mirrors we have the following expressions
similar to \eqref{period}
\be\label{period1}
\sigma(1)_{\vec{\mu}(1)} (\phi_1,...,\phi_{\bar{h}})\sim \sum_{\{m_r\}\in 
\Sigma(1)_{\vec{\mu}(1)}} C(1, \{m_r\})
\prod_{r=1}^{\bar{h}} \frac{\psi_r^{m_r}}{m_r!} \;, 
\ee
where
\be\label{period11}
C(1, \{m_r\})=\prod_{i=1}^5 \Gamma\left(\sum_j(m_r T(1)_{rj}+1) 
B^T(1)_{ji}\right).
\ee
The condition  $\{m_r\}\subset \Sigma_{\vec{\mu}(1)}$  
means that there exist non-negative integers $n_j$, $j=1,...,5$ such that
\be\label{m_r}
m_r T(1)_{rj}=({\vec{\mu}(1)})_j+\sum_k n_{k} M^T(1)_{kj}.
\ee
From \eqref{R-B} it follows that  $ \sum_jT(1)_{rj} B^T(1)_{ji}$, i.e. the factor arising in 
the arguments of the gamma-functions in the formula~\eqref{period11} for the period
$\sigma(1)_{\vec{\mu}(1)}$, coincide with $ \sum_jT(2)_{rj} B^T(2)_{ji}$
in the corresponding expression for the period $\sigma(2)_{\vec{\mu}(2)}$ of the second BHK mirror.\\
The sets of non-negative integers $m_r$ in the expressions for  the periods of both mirrors also coincide, which can be  verified multiplying both parts of the equation \eqref{m_r} by $B(1)_{ji}S_{li}$. Then we obtain the linear system
\be\label{system1}
m_r (\vec{T}(1)_r,\vec{S}_l)=(\mu(1),\vec{S}_l)+ S_{li}n_i.
\ee
Similarly, for the second BHK mirror we get
\be\label{system2}
m_r (\vec{T}(2)_r,\vec{S}_l)=(\mu(2),\vec{S}_l)+ S_{li}n_i.
\ee 
Taking into account the equality~\eqref{SR}, we find that the coefficients 
in both systems of the linear equations ~\eqref{system1} and ~\eqref{system2} are the same.
So that the sets of integers  $m_r$ in the expressions for the periods of both BHK mirrors  also coincide.
Together with the equality~\eqref{R-B}  
it proves the coincidence of  the periods of two BHK mirrors.

\section{Conclusion}
\label{sec:4}
In this paper we showed  that, for any pair of 3-dimensional  Calabi--Yau orbifolds that are multiple BHK mirrors of a loop--chain type pair of Calabi--Yau manifolds in the same weighted projective space, the periods of the holomorphic nonvanishing form coincide.

We assume  that the key observation, leading to this statement, 
 is true  also for the (n-2)-dimensional case. 
  Namely, the following conjecture  is true.
Let $W_1$ and $W_2$ be a loop and a chain polynomials of the form as in (2.4) in the same weighted projective space $\mathbb P(k_1, \ldots , k_n)$.
Then $k_n=1$.
From this conjecture we get that shifted exponents of the monomials of the chiral rings for the loop $W_1$ and the chain $W_2$ 
both contain the basis of the lattice
\[
\begin{matrix}
(-1, 0, 0, \cdots, 0, k_1),\cr
(0, -1, 0, \cdots, 0, k_2),\cr
\cdots \cr
(0, \cdots, 0, -1,  k_{n-1}).\cr
\end{matrix}
\]

It is not clear for us whether the birationality  of the multiple BHK mirrors is equivalent  to the coincidence of their periods, we leave this question for further investigation. 
Even if it is so, we find that the statement about the coincidence  of the multiple BHK mirrors is interesting, in particular taking into account rather straightforward way it is obtained. 

It will be interesting to consider possible generalizations  of this statement  to a wider class of CY varieties.

\section*{Acknowledgments}

The work of A.~Belavin has been supported by the Russian Science Foundation under the grant 18-12-00439.

\newpage

\section*{Appendix.}

 \begin{table}[ht]
\centering
\begin{adjustbox}{width=.89\textwidth}
\small
\begin{tabular}{| r | r | r | r | r | r |}
  \hline
  N & $\{k_i(2)\}$ & $\{A_i\}^{C}\&\{A_i\}^{C(mir)}$ & $\{k_i\}$ & $\{A_i\}^{L}\&\{A_i\}^{L(mir)}$ & $\{k_i(1)\}$ 
  \\ 
  \hline
 1 & \{126,42,70,13,1\} & \{2,3,3,14,239\} & \{92,55,74,17,1\} & \{2,3,3,14,147\} & \{77,26,43,8,1\} \\
 2 & \{270,90,150,13,17\} & \{2,3,3,30,31\} & \{12,7,10,1,1\} & \{2,3,3,30,19\} & \{157,58,91,8,17\} \\
 3 & \{6,2,2,1,1\} & \{2,3,5,10,11\} & \{4,3,2,1,1\} & \{2,3,5,10,7\} & \{83,36,31,16,25\} \\
 4 & \{216,36,66,61,53\} & \{2,6,6,6,7\} & \{3,1,1,1,1\} & \{2,6,6,6,4\} & \{97,25,37,35,53\} \\
 5 & \{64,48,52,51,41\} & \{4,4,4,4,5\} & \{1,1,1,1,1\} & \{4,4,4,4,4\} & \{1,1,1,1,1\} \\
 6 & \{96,32,20,43,1\} & \{2,3,8,4,149\} & \{52,45,14,37,1\} & \{2,3,8,4,97\} & \{62,21,13,28,1\} \\
 7 & \{132,44,20,61,7\} & \{2,3,11,4,29\} & \{10,9,2,7,1\} & \{2,3,11,4,19\} & \{83,30,13,40,7\} \\
 8 & \{160,40,28,73,19\} & \{2,4,10,4,13\} & \{5,3,1,3,1\} & \{2,4,10,4,8\} & \{89,27,17,45,19\} \\
 9 & \{108,27,63,17,1\} & \{2,4,3,9,199\} & \{82,35,59,22,1\} & \{2,4,3,9,117\} & \{63,16,37,10,1\} \\
 10 & \{168,42,98,17,11\} & \{2,4,3,14,29\} & \{12,5,9,2,1\} & \{2,4,3,14,17\} & \{93,26,57,10,11\} \\
 11 &\{324,36,204,37,47\} & \{2,9,3,12,13\} &\{6,1,4,1,1\} & \{2,9,3,12,7\} & \{151,22,109,20,47\} \\
 12 & \{120,24,54,31,11\} & \{2,5,4,6,19\} & \{8,3,4,3,1\} & \{2,5,4,6,11\} & \{64,15,31,18,11\} \\
 13 & \{224,32,104,43,45\} &\{2,7,4,8,9\} & \{4,1,2,1,1\} & \{2,7,4,8,5\} & \{34,7,19,8,15\} \\
 14 & \{180,15,115,49,1\} & \{2,12,3,5,311\} & \{146,19,83,62,1\} & \{2,12,3,5,165\} & \{95,8,61,26,1\} \\
 15 & \{450,15,295,121,19\} & \{2,30,3,5,41\} & \{20,1,11,8,1\} & \{2,30,3,5,21\} & \{221,8,151,62,19\} \\
 16 & \{126,18,78,29,1\} & \{2,7,3,6,223\} & \{100,23,62,37,1\} & \{2,7,3,6,123\} & \{69,10,43,16,1\} \\
 17 & \{324,18,210,73,23\} & \{2,18,3,6,25\} & \{12,1,7,4,1\} & \{2,18,3,6,13\} & \{157,10,109,38,23\} \\
 18 & \{288,24,184,49,31\} & \{2,12,3,8,17\} & \{8,1,5,2,1\} & \{2,12,3,8,9\} & \{137,14,97,26,31\} \\
 19 & \{112,16,52,43,1\} & \{2,7,4,4,181\} & \{80,21,34,45,1\} & \{2,7,4,4,101\} & \{62,9,29,24,1\} \\
 20 & \{128,16,60,49,3\} & \{2,8,4,4,69\} & \{31,7,13,17,1\} & \{2,8,4,4,38\} & \{23,3,11,9,1\} \\
 21 & \{160,16,76,61,7\} & \{2,10,4,4,37\} & \{17,3,7,9,1\} & \{2,10,4,4,20\} & \{83,9,41,33,7\} \\
 22 & \{272,16,132,103,21\} & \{2,17,4,4,21\} & \{10,1,4,5,1\} & \{2,17,4,4,11\} & \{44,3,23,18,7\} \\
 23 & \{180,45,21,113,1\} & \{2,4,15,3,247\} & \{94,59,11,82,1\} & \{2,4,15,3,153\} & \{111,28,13,70,1\} \\
 24 & \{300,75,21,193,11\} & \{2,4,25,3,37\} & \{14,9,1,12,1\} & \{2,4,25,3,23\} & \{181,48,13,120,11\} \\
 25 & \{234,39,33,145,17\} & \{2,6,13,3,19\} & \{8,3,1,6,1\} & \{2,6,13,3,11\} & \{127,24,19,84,17\} \\
 26 & \{100,20,36,41,3\} & \{2,5,5,4,53\} & \{22,9,8,13,1\} & \{2,5,5,4,31\} & \{19,4,7,8,1\} \\
 27 & \{210,15,81,113,1\} & \{2,14,5,3,307\} & \{144,19,41,102,1\} & \{2,14,5,3,163\} & \{111,8,43,60,1\} \\
 28 & \{480,15,189,257,19\} & \{2,32,5,3,37\} & \{18,1,5,12,1\} & \{2,32,5,3,19\} & \{237,8,97,132,19\} \\
 29 & \{324,27,69,193,35\} & \{2,12,9,3,13\} & \{6,1,1,4,1\} & \{2,12,9,3,7\} & \{157,16,37,104,35\} \\
 30 & \{220,20,84,89,27\} & \{2,11,5,4,13\} & \{6,1,2,3,1\} & \{2,11,5,4,7\} & \{35,4,15,16,9\} \\
 31 & \{224,28,60,97,39\} & \{2,8,7,4,9\} & \{4,1,1,2,1\} & \{2,8,7,4,5\} & \{35,6,11,18,13\} \\
 32 & \{24,24,16,7,1\} & \{3,2,3,8,65\} & \{14,23,19,8,1\} & \{3,2,3,8,51\} & \{37,38,25,11,2\} \\
 33 & \{27,27,18,7,2\} & \{3,2,3,9,37\} & \{8,13,11,4,1\} & \{3,2,3,9,29\} & \{41,43,28,11,4\} \\
 34 & \{33,33,22,7,4\} & \{3,2,3,11,23\} & \{5,8,7,2,1\} & \{3,2,3,11,18\} & \{49,53,34,11,8\} \\
 35 & \{25,25,10,13,2\} & \{3,2,5,5,31\} & \{6,13,5,6,1\} & \{3,2,5,5,25\} & \{39,41,16,21,4\} \\
 36 & \{32,48,20,27,1\} & \{4,2,4,4,101\} & \{15,41,19,25,1\} & \{4,2,4,4,86\} & \{27,41,17,23,1\} \\
 37 & \{72,108,30,43,35\} & \{4,2,6,6,7\} & \{1,3,1,1,1\} & \{4,2,6,6,6\} & \{53,97,25,37,35\} \\
 38 & \{28,28,8,19,1\} & \{3,2,7,4,65\} & \{12,29,7,16,1\} & \{3,2,7,4,53\} & \{45,46,13,31,2\} \\
 39 & \{52,52,8,37,7\} & \{3,2,13,4,17\} & \{3,8,1,4,1\} & \{3,2,13,4,14\} & \{81,88,13,61,14\} \\
 40 & \{9,9,2,5,2\} & \{3,2,9,5,11\} & \{2,5,1,2,1\} & \{3,2,9,5,9\} & \{67,77,16,41,20\} \\
 41 & \{30,45,25,19,1\} & \{4,2,3,5,101\} & \{16,37,27,20,1\} & \{4,2,3,5,85\} & \{25,38,21,16,1\} \\
 42 & \{36,54,30,19,5\} & \{4,2,3,6,25\} & \{4,9,7,4,1\} & \{4,2,3,6,21\} & \{29,46,25,16,5\} \\
 43 & \{18,99,39,59,1\} & \{12,2,3,3,157\} & \{8,61,35,52,1\} & \{12,2,3,3,149\} & \{17,94,37,56,1\} \\
 44 & \{9,54,21,32,1\} & \{13,2,3,3,85\} & \{4,33,19,28,1\} & \{13,2,3,3,81\} & \{17,103,40,61,2\} \\
 45 & \{9,63,24,37,2\} & \{15,2,3,3,49\} & \{2,19,11,16,1\} & \{15,2,3,3,47\} & \{17,121,46,71,4\} \\
 46 & \{9,81,30,47,4\} & \{19,2,3,3,31\} & \{1,12,7,10,1\} & \{19,2,3,3,30\} & \{17,157,58,91,8\} \\
 47 & \{15,45,20,17,8\} & \{7,2,3,5,11\} & \{1,4,3,2,1\} & \{7,2,3,5,10\} & \{25,83,36,31,16\} \\
 48 & \{66,99,15,83,1\} & \{4,2,11,3,181\} & \{24,85,11,60,1\} & \{4,2,11,3,157\} & \{57,86,13,72,1\} \\
 49 & \{126,189,15,163,11\} & \{4,2,21,3,31\} & \{4,15,1,10,1\} & \{4,2,21,3,27\} & \{107,166,13,142,11\} \\
 50 & \{21,42,9,32,1\} & \{5,2,7,3,73\} & \{8,33,7,24,1\} & \{5,2,7,3,65\} & \{37,75,16,57,2\} \\
       51 & \{39,78,9,62,7\} & \{5,2,13,3,19\} & \{2,9,1,6,1\} & \{5,2,13,3,17\} & \{67,141,16,111,14\} \\
 52 & \{27,81,12,59,10\} & \{7,2,9,3,13\} & \{1,6,1,4,1\} & \{7,2,9,3,12\} & \{47,151,22,109,20\} \\
  53 & \{7,14,3,8,3\} & \{5,2,7,4,9\} & \{1,4,1,2,1\} & \{5,2,7,4,8\} & \{15,34,7,19,8\} \\
 54 & \{30,75,21,53,1\} & \{6,2,5,3,127\} & \{12,55,17,42,1\} & \{6,2,5,3,115\} & \{27,68,19,48,1\} \\
   55 & \{15,45,12,31,2\} & \{7,2,5,3,37\} & \{3,16,5,12,1\} & \{7,2,5,3,34\} & \{27,83,22,57,4\} \\
 56 & \{72,48,84,11,1\} & \{3,3,2,12,205\} & \{56,37,94,17,1\} & \{3,3,2,12,149\} & \{52,35,61,8,1\} \\
 57 & \{168,112,196,11,17\} & \{3,3,2,28,29\} & \{8,5,14,1,1\} & \{3,3,2,28,21\} & \{116,83,141,8,17\} \\
    58 & \{60,24,78,17,1\} & \{3,5,2,6,163\} & \{48,19,68,27,1\} & \{3,5,2,6,115\} & \{42,17,55,12,1\} \\
   59 & \{168,48,228,23,37\} & \{3,7,2,12,13\} & \{4,1,6,1,1\} & \{3,7,2,12,9\} & \{104,35,157,16,37\} \\
 60 & \{32,24,52,19,1\} & \{4,4,2,4,109\} & \{23,17,41,27,1\} & \{4,4,2,4,86\} & \{25,19,41,15,1\} \\
 \hline
\end{tabular}
\end{adjustbox}
\end{table}

 \begin{table}[ht]
\centering
\begin{adjustbox}{width=.89\textwidth}
\small
\begin{tabular}{| r | r | r | r | r | r | }
  \hline
  N & $\{k_i(2)\}$ & $\{A_i\}^{C}\&\{A_i\}^{C(mir)}$ & $\{k_i\}$ & $\{A_i\}^{L}\&\{A_i\}^{L(mir)}$ & $\{k_i(1)\}$ 
  \\ 
  \hline
 61 & \{80,48,136,23,33\} & \{4,5,2,8,9\} & \{2,1,4,1,1\} & \{4,5,2,8,7\} & \{18,13,35,6,11\} \\
 62 & \{88,16,124,35,1\} & \{3,11,2,4,229\} & \{72,13,86,57,1\} & \{3,11,2,4,157\} & \{60,11,85,24,1\} \\
 63 & \{184,16,268,71,13\} & \{3,23,2,4,37\} & \{12,1,14,9,1\} & \{3,23,2,4,25\} & \{120,11,181,48,13\} \\
 64 & \{132,24,186,35,19\} & \{3,11,2,6,19\} & \{6,1,8,3,1\} & \{3,11,2,6,13\} & \{84,17,127,24,19\} \\
 65 & \{64,24,116,35,17\} & \{4,8,2,4,13\} & \{3,1,5,3,1\} & \{4,8,2,4,10\} & \{45,19,89,27,17\} \\
 66 & \{48,60,114,29,37\} & \{6,4,2,6,7\} & \{1,1,3,1,1\} & \{6,4,2,6,6\} & \{35,53,97,25,37\} \\
 67 & \{24,32,44,19,1\} & \{5,3,2,4,101\} & \{16,21,38,25,1\} & \{5,3,2,4,85\} & \{20,27,37,16,1\} \\
 68 & \{140,56,26,197,1\} & \{3,5,14,2,223\} & \{60,43,8,111,1\} & \{3,5,14,2,163\} & \{102,41,19,144,1\} \\
 69 & \{75,30,13,106,1\} & \{3,5,15,2,119\} & \{32,23,4,59,1\} & \{3,5,15,2,87\} & \{109,44,19,155,2\} \\
 70 & \{85,34,13,121,2\} & \{3,5,17,2,67\} & \{18,13,2,33,1\} & \{3,5,17,2,49\} & \{123,50,19,177,4\} \\
 71 & \{105,42,13,151,4\} & \{3,5,21,2,41\} & \{11,8,1,20,1\} & \{3,5,21,2,30\} & \{151,62,19,221,8\} \\
 72 & \{54,18,16,73,1\} & \{3,6,9,2,89\} & \{25,14,5,44,1\} & \{3,6,9,2,64\} & \{77,26,23,105,2\} \\
 73 & \{78,26,16,109,5\} & \{3,6,13,2,25\} & \{7,4,1,12,1\} & \{3,6,13,2,18\} & \{109,38,23,157,10\} \\
 74 & \{49,14,19,64,1\} & \{3,7,7,2,83\} & \{24,11,6,41,1\} & \{3,7,7,2,59\} & \{69,20,27,91,2\} \\
 75 & \{72,18,22,97,7\} & \{3,8,9,2,17\} & \{5,2,1,8,1\} & \{3,8,9,2,12\} & \{97,26,31,137,14\} \\
 76 & \{56,42,26,99,1\} & \{4,4,7,2,125\} & \{24,29,9,62,1\} & \{4,4,7,2,101\} & \{45,34,21,80,1\} \\
 77 & \{64,48,26,115,3\} & \{4,4,8,2,47\} & \{9,11,3,23,1\} & \{4,4,8,2,38\} & \{17,13,7,31,1\} \\
 78 & \{88,66,26,163,9\} & \{4,4,11,2,21\} & \{4,5,1,10,1\} & \{4,4,11,2,17\} & \{23,18,7,44,3\} \\
 79 & \{50,30,34,83,3\} & \{4,5,5,2,39\} & \{8,7,4,19,1\} & \{4,5,5,2,31\} & \{13,8,9,22,1\} \\
 80 & \{120,16,86,137,1\} & \{3,15,4,2,223\} & \{70,13,28,111,1\} & \{3,15,4,2,153\} & \{82,11,59,94,1\} \\
 81 & \{216,16,158,245,13\} & \{3,27,4,2,31\} & \{10,1,4,15,1\} & \{3,27,4,2,21\} & \{142,11,107,166,13\} \\
 82 & \{50,10,28,61,1\} & \{3,10,5,2,89\} & \{27,8,9,44,1\} & \{3,10,5,2,62\} & \{69,14,39,85,2\} \\
 83 & \{55,10,31,67,2\} & \{3,11,5,2,49\} & \{15,4,5,24,1\} & \{3,11,5,2,34\} & \{75,14,43,93,4\} \\
 84 & \{65,10,37,79,4\} & \{3,13,5,2,29\} & \{9,2,3,14,1\} & \{3,13,5,2,20\} & \{87,14,51,109,8\} \\
 85 & \{85,10,49,103,8\} & \{3,17,5,2,19\} & \{6,1,2,9,1\} & \{3,17,5,2,13\} & \{111,14,67,141,16\} \\
 86 & \{84,14,34,109,11\} & \{3,12,7,2,13\} & \{4,1,1,6,1\} & \{3,12,7,2,9\} & \{109,20,47,151,22\} \\
 87 & \{80,30,58,131,21\} & \{4,8,5,2,9\} & \{2,1,1,4,1\} & \{4,8,5,2,7\} & \{19,8,15,34,7\} \\
 88 & \{72,96,22,169,1\} & \{5,3,12,2,191\} & \{26,61,8,95,1\} & \{5,3,12,2,165\} & \{62,83,19,146,1\} \\
 89 & \{39,52,11,92,1\} & \{5,3,13,2,103\} & \{14,33,4,51,1\} & \{5,3,13,2,89\} & \{67,90,19,159,2\} \\
 90 & \{45,60,11,107,2\} & \{5,3,15,2,59\} & \{8,19,2,29,1\} & \{5,3,15,2,51\} & \{77,104,19,185,4\} \\
 91 & \{57,76,11,137,4\} & \{5,3,19,2,37\} & \{5,12,1,18,1\} & \{5,3,19,2,32\} & \{97,132,19,237,8\} \\
 92 & \{42,70,26,113,1\} & \{6,3,7,2,139\} & \{16,43,10,69,1\} & \{6,3,7,2,123\} & \{37,62,23,100,1\} \\
 93 & \{54,90,26,149,5\} & \{6,3,9,2,35\} & \{4,11,2,17,1\} & \{6,3,9,2,31\} & \{47,80,23,132,5\} \\
 94 & \{14,14,8,31,3\} & \{5,4,7,2,13\} & \{2,3,1,6,1\} & \{5,4,7,2,11\} & \{15,16,9,35,4\} \\
 95 & \{18,78,58,97,1\} & \{14,3,3,2,155\} & \{8,43,26,77,1\} & \{14,3,3,2,147\} & \{17,74,55,92,1\} \\
 96 & \{9,42,31,52,1\} & \{15,3,3,2,83\} & \{4,23,14,41,1\} & \{15,3,3,2,79\} & \{17,80,59,99,2\} \\
 97 & \{9,48,35,59,2\} & \{17,3,3,2,47\} & \{2,13,8,23,1\} & \{17,3,3,2,45\} & \{17,92,67,113,4\} \\
 98 & \{9,60,43,73,4\} & \{21,3,3,2,29\} & \{1,8,5,14,1\} & \{21,3,3,2,28\} & \{17,116,83,141,8\} \\
 99 & \{24,64,38,89,1\} & \{9,3,4,2,127\} & \{10,37,16,63,1\} & \{9,3,4,2,117\} & \{22,59,35,82,1\} \\
 100 & \{24,80,46,109,5\} & \{11,3,4,2,31\} & \{2,9,4,15,1\} & \{11,3,4,2,29\} & \{22,75,43,102,5\} \\
 101 & \{36,84,34,127,7\} & \{8,3,6,2,23\} & \{2,7,2,11,1\} & \{8,3,6,2,21\} & \{32,77,31,116,7\} \\
 102 & \{21,56,19,85,8\} & \{9,3,7,2,13\} & \{1,4,1,6,1\} & \{9,3,7,2,12\} & \{37,104,35,157,16\} \\
 103 & \{32,40,38,77,5\} & \{6,4,4,2,23\} & \{3,5,3,11,1\} & \{6,4,4,2,20\} & \{27,35,33,67,5\} \\
 104 & \{20,30,22,59,9\} & \{7,4,5,2,9\} & \{1,2,1,4,1\} & \{7,4,5,2,8\} & \{11,18,13,35,6\} \\
 105 & \{48,18,58,67,1\} & \{4,8,3,2,125\} & \{28,13,21,62,1\} & \{4,8,3,2,97\} & \{37,14,45,52,1\} \\
 106 & \{84,18,106,115,13\} & \{4,14,3,2,17\} & \{4,1,3,8,1\} & \{4,14,3,2,13\} & \{61,14,81,88,13\} \\
 107 & \{9,6,13,16,1\} & \{5,6,3,2,29\} & \{5,4,5,14,1\} & \{5,6,3,2,24\} & \{29,20,43,53,4\} \\
 108 & \{21,12,31,37,4\} & \{5,7,3,2,17\} & \{3,2,3,8,1\} & \{5,7,3,2,14\} & \{33,20,51,61,8\} \\
 109 & \{27,12,41,47,8\} & \{5,9,3,2,11\} & \{2,1,2,5,1\} & \{5,9,3,2,9\} & \{41,20,67,77,16\} \\
 110 & \{48,40,62,113,25\} & \{6,6,4,2,7\} & \{1,1,1,3,1\} & \{6,6,4,2,6\} & \{37,35,53,97,25\} \\
 111 & \{12,18,22,31,1\} & \{7,4,3,2,53\} & \{6,11,9,26,1\} & \{7,4,3,2,47\} & \{21,32,39,55,2\} \\
 \hline
\end{tabular}
\end{adjustbox}
\caption{
Here $\{k_i\}$ are weights arising simultaneously in loop and chain types, with the corresponding exponents $\{A_i\}^C$ and $\{A_i\}^L$ in~\eqref{A_i}. 
The $\{k_i(2)\}$ and $\{A_i\}^{C(mir)}$ are the weights and the exponents of the Mirror CY for the chain manifold $\{A_i\}^C$. The  sets $\{k_i(1)\}$ and $\{A_i\}^{L(mir)}$ are the same for the loop manifold $\{A_i\}^L$. 
The notation $\{A_i\}^{C}\&\{A_i\}^{C(mir)}$ means that two sets coincide.
 }
\label{table:Chain-Loop-Mirrors}
\end{table}

\end{document}